\documentclass[12pt]{article}

\pdfoutput=1

\usepackage[top=80pt,bottom=85pt,left=85pt,right=85pt]{geometry}
\usepackage{amssymb}
\usepackage{amsmath}
\usepackage{setspace}
\usepackage{sectsty}
\usepackage{cite}
\usepackage{bbm}
\usepackage[utf8]{inputenc}
\usepackage[T1]{fontenc}
\usepackage{comment}
\usepackage[english]{babel}
\usepackage{afterpage}
\usepackage{bbold}

\usepackage[usenames]{xcolor}
\usepackage{graphicx,subcaption}
\usepackage{setspace}
\usepackage{float}
\usepackage{booktabs}
\usepackage[debug,pageanchor=false]{hyperref}
\definecolor{link}{rgb}{.8,.15,.1}
\definecolor{pigment}{rgb}{0.36, 0.54, 0.66}
\definecolor{pigment2}{rgb}{0.19, 0.55, 0.91}
\definecolor{pigment3}{rgb}{0.2, 0.2, 0.6}
\definecolor{light-gray}{gray}{0.75}
\hypersetup{colorlinks=true,linkcolor=link,citecolor=link,urlcolor=link,linktocpage}

\usepackage{array,multirow,booktabs,longtable}
\usepackage{mathrsfs}
\usepackage{tikz-cd} 
\usetikzlibrary{backgrounds, arrows,calc,shapes,decorations.pathreplacing, decorations.markings, automata,positioning}
\tikzset{%
  >={Latex[width=2mm,length=2mm]},
            base/.style = {rectangle, rounded corners, draw=black,
                           minimum width=4cm, minimum heigwht=1cm,
                           text centered, font=\sffamily},
  activityStarts/.style = {base, fill=orange!15},
       startstop/.style = {base, fill=orange!15},
    activityRuns/.style = {base, fill=orange!15},
         process/.style = {base, minimum width=2.5cm, fill=orange!15,
                           font=\ttfamily},
}
\newcommand{\red}[1]{}
\tikzset{
        cvertex/.style={circle,draw=black,inner sep=1pt,outer sep=3pt},
        vertex/.style={circle,fill=black,inner sep=1pt,outer sep=3pt},
        star/.style={circle,fill=yellow,inner sep=0.75pt,outer sep=0.75pt},
        tvertex/.style={inner sep=1pt,font=\scriptsize},
        gap/.style={inner sep=0.5pt,fill=white}}

\tikzstyle{mybox} = [draw=black, fill=blue!10, very thick,
    rectangle, rounded corners, inner sep=10pt, inner ysep=20pt]
\tikzstyle{boxtitle} =[fill=blue!50, text=white,rectangle,rounded corners]

\newcommand{\cc}{\mathbb{C}}
\newcommand{\zz}{\mathbb{Z}}
\newcommand{\pp}{\mathbb{P}}

\def\cO{\mathcal{O}}

\newcommand{\todo}[1]{}
\renewcommand{\todo}[1]{{\color{red} TODO: {#1}}}
\renewcommand{\red}[1]{{\color{red} {#1}}}

\newcommand{\be}{\begin{equation}}  
\newcommand{\ee}{\end{equation}}  
\newcommand{\bea}{\begin{align}}
\newcommand{\eea}{\end{align}}
\newcommand{\bp}{\begin{bmatrix*}[r]}  
\newcommand{\ep}{\end{bmatrix*}}  
\newcommand{\bpp}{\begin{bmatrix}}  
\newcommand{\epp}{\end{bmatrix}}  
\newcommand{\bcd}{\begin{center}
\begin{tikzcd}}
\newcommand{\ecd}{\end{tikzcd} \end{center}}
\newcommand{\bm}{\begin{pmatrix}}  
\newcommand{\eem}{\end{pmatrix}}

\makeatletter
\@addtoreset{equation}{section}
\makeatother
\begin{document}

% titlepage 

\begin{titlepage}

\begin{center}

\vskip .3in \noindent

{\Large \bf{Higgs branches of 5d rank-zero theories \\  from geometry}}

\bigskip\bigskip

Andr\'es Collinucci$^a$, Mario De Marco$^{b}$, Andrea Sangiovanni$^{c}$ and Roberto Valandro$^{c}$ \\

\bigskip

%Version of \DTMNow~(w.r.t. GMT)

\bigskip
{\footnotesize
 \it

$^a$ Service de Physique Th\'eorique et Math\'ematique, Universit\'e Libre de Bruxelles and \\ International Solvay Institutes, Campus Plaine C.P.~231, B-1050 Bruxelles, Belgium\\
\vspace{.25cm}
$^b$ SISSA and INFN, Via Bonomea 265, I-34136 Trieste, Italy\\
\vspace{.25cm}
$^c$ Dipartimento di Fisica, Universit\`a di Trieste, Strada Costiera 11, I-34151 Trieste, Italy \\%and \\
%\vspace{.25cm}
and INFN, Sezione di Trieste, Via Valerio 2, I-34127 Trieste, Italy	
}

\vskip .5cm
{\scriptsize \tt collinucci dot phys at gmail dot com \hspace{2cm}  mdemarco at sissa dot it} \\
{\scriptsize \tt \hspace{0.5cm}  andrea dot sangiovanni at phd dot units dot it \hspace{1cm}  roberto dot valandro at ts dot infn dot it}

\vskip 1cm
     	{\bf Abstract }
\vskip .1in
\end{center}
We study the Higgs branches of five-dimensional $\mathcal{N}=1$ rank-zero theories obtained from M-theory on two classes non-toric non-compact Calabi-Yau threefolds: Reid's pagodas, and Laufer's examples. Our approach consists in reducing to IIA with D6-branes and O6-planes, and computing the open-string spectra giving rise to hypermultiplets.
Starting with the seven-dimensional worldvolume theories, we switch on T-brane backgrounds to give rise to bound states with angles. We observe that the resulting partially Higgsed 5d theories have discrete gauge groups, from which we readily deduce the geometry of the Higgs branches as orbifolds of quaternionic varieties.

\noindent

\vfill
\eject

\end{titlepage}

% end titlepage

\tableofcontents

%\newpage 
%%%%%%%%%%%%%%%%%%%%%%%%%%%
\section{Introduction} % sec (intro)
\label{sec:intro}
%%%%%%%%%%%%%%%%%%%%%%%%%%%
Higgs branches of theories with eight supercharges are interesting and highly adaptive geometric objects to study. Interesting, because they are immune to quantum corrections\footnote{Some caveats apply, such as the possibility of dimension jumps when a gauge coupling is strictly infinite, see \cite{Ferlito:2017xdq} for a seminar reference on the subject. Also, the possibility of quantum splitting can spoil this picture, we thank Julius Grimminger for pointing out counterexamples to this creed, \cite{Grimminger:2020dmg, Assel:2017jgo}.}. Adaptive, because they do not change under dimensional reduction, which means that one can study them in any dimension compatible with eight supercharges.

Having a Lagrangian formulation of a theory allows one to at least in principle construct its Higgs branch as a hyper-K\"ahler quotient. Often, however, this is not possible to do directly. For instance, a 5d $\mathcal{N}=1$ Lagrangian theory may have a UV fixed point that is infinitely strongly coupled. In that case, one cannot trust the construction at weak coupling. This issue has been studied in many papers, such as \cite{Ferlito:2017xdq}.

M-theory on non-compact singular Calabi-Yau threefolds is a well-known arena for constructing five-dimensional SCFT's. This subject was initiated in the nineties in the two seminal papers \cite{Intriligator:1997pq,Seiberg:1996bd}. 
In the past five years, the subject has seen a revival, with more thorough systematic studies of such theories, their global symmetries, their moduli spaces, their prepotentials, and various methods for constructing them. It is difficult at this point to reference all works, some key works include \cite{Closset:2020afy,vanBeest:2020civ,Apruzzi:2019vpe,Bhardwaj:2018vuu,Collinucci:2008pf,Apruzzi:2019opn,Apruzzi:2019enx, Bhardwaj:2019jtr}.
From that perspective, the Higgs branch will be realized by the complex structure moduli and the Wilson lines of the supergravity $C_3$-form. 

Mostly, the literature focuses on toric threefolds. Those are particularly nice for several reasons: They can be constructed by drawing a two-dimensional diagram, the effective field theory data can be read off such a diagram, the prepotential can be understood in terms of the topological vertex, and one can directly relate it to 5-brane webs in IIB via a chain of dualities, \cite{Aharony:1997bh}.

However, the toric realm is but a tiny subset of all possible interesting geometries one can build. Indeed, even if we just restrict to the hypersurface case, the toric realm misses the majority of situations. It is the purpose of this paper to initiate the study of non-toric affine threefolds. We will focus on the case of rank-zero theories, and derive Higgs branches for several classes of threefolds. We will find that they behave like free hypers with discrete gauging. 

We will focus on threefolds that admit $\cc^*$-fibrations, allowing for reductions to IIA string theory with D6-branes and O6-planes. This class of examples admits a straightforward relation between IIA and M-theory as follows. Given a stack of $N$ D6-branes, there are three adjoint scalars. Pairing up two of them into a complex Higgs field $\Phi$, the relation between IIA and M-theory is given by the following:
\be
u v = \det \big( z \cdot \mathbb{1}_N-\langle \Phi \rangle \big)\,.
\ee
So the M-theory geometry follows the spectral data of the worldvolume Higgs field. With this viewpoint, the Higgs branch of the D6-theory, which is defined by the fluctuations of $\Phi$, immediately translates into geometric data in M-theory.

In this paper, we will study threefolds that are one-parameter families of A-type and D-type that admit simple flops. `Simple flops' means that these threefolds admit small resolutions such that only one exceptional $\pp^1$ is produced. 
We recast the problem of parametrizing complex structure deformations in terms of switching on open string vev's in IIA string theory with D6-branes and O6-planes.

For the A-series, we will tackle the so-called family of Reid's \emph{pagodas}. These are a class of conifold-like threefolds, which, despite being simple flops, can have Higgs branches of arbitrarily high dimensions. We will fiberwise reduce M-theory on such threefolds to IIA on $\cc^2$ with intersecting D6-branes, and capture these hypermultiplets through Ext$^1$ computations of open string spectra. We will confirm and extend the results of \cite{Closset:2020scj}, and will find that the Higgs branches are given by
\begin{equation}
\mathcal{M}_{\text{RP}_k} = \mathbb{H}^k/\zz_k\,,
\end{equation}
where $k \in \zz$ is an integer parametrizing the family of Reid's Pagoda threefolds. The flavor group of the theory turns out to be $G_{\rm fl} = U(1)/\zz_k$.
The presence of a single Abelian factor means that only one real mass can be switched on, which matches the fact that Reid's pagodas only admit simple flops.

For the D-series, we will study two families of threefolds: The so-called Brown-Wemyss family \cite{BrownWemyss}, and the family of \emph{Laufer's examples} \cite{laufer}. These are the so-called `flops of length two'. The `length two' here means that, even though the threefold admits only a simple flop, the exceptional $\pp^1$ will allow for bound states of two M2-branes to wrap it, a curious property conferred by the $\cO(-3) \oplus \cO(1)$ normal bundle. We will recast this information in terms of IIA string theory with D6-branes and O6-planes. The orientifolding will allow for charge-two open string states. We will find that the Higgs branch for the Brown-Wemyss case is given by
\begin{equation}
\mathcal{M}_{\rm HB} = \mathbb{H} \times \mathbb{H}^4/\mathbb{Z}_2\times \mathbb{H}/\mathbb{Z}_2     
\end{equation}
whereas for Laufer's examples one has
\begin{equation}
\mathcal{M}_{\rm HB} =  \mathbb{H}^k \times \mathbb{H}^{2k+2}/\mathbb{Z}_2\times \mathbb{H}/\mathbb{Z}_2     
\end{equation}
where $k \in \zz$ is an integer parametrizing the family of Laufer's threefolds. Here the flavor group is $G_F=U(1)$. Again, this matches the fact that these spaces only admit simple flops.

\section{The conifold}
In this section, we will introduce the simplest possible example: the Higgs branch of a free hypermultiplet, seen in terms of the moduli space of M-theory on the conifold.
We define the conifold as the following hypersurface in $\cc^4$:
\be\label{Eq:conifold}
u v = z^2-w^2 \, \quad \subset \quad \cc^4 \langle u, v, z, w \rangle\,.
\ee
This space admits two inequivalent small resolutions:
\begin{align}
\bm u & z-w\\z+w & v \eem \cdot \bm s \\t \eem &= 0 \quad {\rm and} \qquad
\bm u & z+w\\z-w & v \eem \cdot \bm s \\t \eem = 0\,,
\end{align}
where $[s:t]$ are the homogeneous coordinates of a $\pp^1$. These small resolutions replace the singularity with a $\pp^1$, on which an M2-brane can be wrapped. This sphere has a normal bundle $\cO(-1) \oplus \cO(-1)$. From the five-dimensional perspective, this implies that this M2-brane will correspond to a hypermultiplet, as explained by Witten \cite{Witten:1996qb}.

This conifold admits one normalizable deformation
\be\label{normalizable}
u v = z^2-w^2+\mu\,, \quad {\rm with} \quad \mu \in \cc\,.
\ee
This mode sits in $H^3(X_3, \cc)$, where $X_3$ is the threefold defined in (\ref{normalizable}). After deforming, a 3-sphere $A$ is created, and the dual non-compact 3-cycle $B$ intersects $A$ at a point. In terms of the holomorphic 3-form $\Omega_3$, we can think of the value of $\mu$ as the period
\be
\mu: = \int_{A} \Omega_3\,.
\ee
In terms of Poincar\'e dual 3-forms $\alpha$ and $\beta$, we have that
\be
\int_{X_3} \alpha \wedge \beta = 1
\ee
The supergravity $C_3$-field can thus be `dimensionally reduced' along these forms as
\be
C_3 = a \alpha + b \beta\,,
\ee
where $(a, b)$ form a pair of real Wilson lines for the 3-form. We can now combine all four real degrees of freedom $(\mu, a, b)$ into a single five-dimensional hypermultiplet. The Higgs branch for this theory is a single-centered Taub-Nut space, of quaternionic dimension $one$:
\be
\mathcal{H} = {\rm TN}_1 \cong \mathbb{H}\,.
\ee
In this paper, we will study such Higgs branches by exploiting the duality between M-theory on $\cc^*$-fibered threefolds and IIA string theory in the presence of D6-branes. In this case, the $u v = \ldots$ form of our hypersurface shows us that the conifold is indeed $\cc^*$-fibered, and that the circle in $\cc^*  \cong \mathbb{R} \times S^1$ collapses wherever the r.h.s. vanishes, i.e. over the reducible locus
\be
{\rm D6-locus}: \quad (z-w) (z+w)=0\,.
\ee
The link between M-theory geometry and IIA D6-configurations is established as follows: Let there be a stack of $N$ coincident D6-branes in $\cc^2 \langle z, w \rangle$, located on the divisor $z=0$. This stack gives rise to $\mathcal{N}=1, d=7$ SYM with $SU(N)$ gauge group. There are three real adjoint scalars in the vector multiplet: $\phi_{i=1, 2, 3}$. One pairs up two of these into a complex field $\Phi := \phi_1 + i \phi_2$. In this case, we begin with two parallel D6-branes, at the origin of the Higgs branch, i.e. $\langle \Phi \rangle=0$. Then, the M-theory uplift is given by a $\cc^*$-fibration that collapses over the spectral curve of $\Phi$ as follows:
\be
u v = \det \big( z \mathbb{1}_2-\langle \Phi \rangle \big) = z^2\,.
\ee
In this case, the geometry is that of a local K3 with an $A_1$-singularity times a complex plane generated by $w$. This describes 7d $\mathcal{N}=1$ SYM. Now consider switching on the following position-dependent vev
\be
\langle \Phi \rangle = \bm w & 0 \\ 0 & -w \eem\,.
\ee
This breaks $SU(2) \mapsto U(1)$, although the group enhances back to $SU(2)$ at the origin $(w,z)=(0,0)$. Now we have two intersecting branes, and the M-theory geometry is given by the threefold:
\be
u v = z^2-w^2 \quad \subset \quad \cc^4\,.
\ee
Supersymmetry is broken to eight supercharges. We expect there to be a free 5d hypermultiplet. In order to see this in IIA, we use the linearized equations of motion in holomorphic gauge for the fluctuation field $\varphi$ of the Higgs, as explained in the seven-brane case in \cite{Cecotti:2010bp}:
\be
\partial \varphi = 0 \qquad \varphi \sim \varphi + [\langle \Phi \rangle, g]\,,
\ee
where $g$ are complexified $\mathfrak{su}(2)$ matrices, i.e. $g\in\mathfrak{sl}(2)$. Parametrizing both the fluctuation and gauge parameter as follows:
\be
\varphi= \bm \varphi_0 & \varphi_+\\ \varphi_- & -\varphi_0 \eem\,, \qquad g = \tfrac{1}{2}\,\bm g_0& g_+\\g_-& -g_0 \eem\,,
\ee
we deduce that
\be
\varphi \sim \varphi + w \bm 0&g_+\\-g_-&0 \eem\,.
\ee
This tells us a few things. First, that we can have 7d fluctuations given by $\varphi_0$. Most importantly, that the fluctuations $\varphi_\pm$ are defined up to any multiple of $w$. This means that they are localized on the $(w, z)=(0,0)$ locus, and are therefore genuinely 5d dynamical fields:
\be
\varphi_\pm \in \cc[w]/(w) \cong \cc\,.
\ee
The pair $(\varphi_+, \varphi_-)$ forms a free hypermultiplet, as expected. If we switch on a vev for this pair, the M-theory geometry deforms as follows:
\be\label{single brane}
u v = z^2-w^2 -\varphi_+ \varphi_-\,.
\ee
Therefore, we see that there is a projection from the full hypermultiplet moduli space onto the complex structure moduli space of the conifold
\be
\pi: (\varphi_+, \varphi_-)  \longrightarrow \mu:=\varphi_+ \varphi_-\,.
\ee
This map defines the Higgs branch as a $\cc^*$-fibration over the complex structure moduli space, whereby the fibers contain the data about the $C_3$ Wilson lines. From the brane perspective this is understood from the fact that there is an action
\be
 (\varphi_+, \varphi_-) \mapsto  (\lambda \varphi_+, \lambda^{-1}\varphi_-)\,,
\ee
where $\lambda$ is a parameter in the complexified flavor group $U(1)_\cc \cong \cc^*$. So, if we recombine the two D6-branes, we will get a single brane in the shape of a throat (diffeomorphic to a cylinder), as it can be seen from the right hand side of (\ref{single brane}). One can then define two Wilson lines for the worldvolume gauge field: one along the compact 1-cycle, and one along the non-compact 1-cycle.  This uplifts in M-theory to the two Wilson lines of the supergravity $C_3$ form on the $S^3$, and its dual non-compact 3-cycle.
 
\section{ADE families and IIA Higgs field}\label{Sec:ADEfamilies}

The conifold presented in the previous section is a particular example of a family of A-type ALE spaces, parametrized by the parameter $w$. At $w=0$ the  equation \eqref{Eq:conifold} describes an ALE surface with a $A_1$ singularity at the origin; at generic value of $w$ the singularity is deformed. 

Let us generalize it to an $A_{n-1}$ type ALE family. This machinery is well-developed in \cite{Katz:1992aa}. The $A_{n-1}$ singularity has the form
\be 
 uv=z^n \:.
\ee
Its versal deformation is
\be \label{Anfamily}
 uv=z^n + \sum _{i=2}^n (-1)^i\sigma_i z^{n-i}\:.
\ee
This space is a fibration with fiber given by the (deformed) $A_{n-1}$ surface and  base the space $\mathfrak{t}/\mathcal{W}$ of gauge invariant coordinates $\sigma_i$ ($i=2,...,n$)\footnote{The invariant $\sigma_i$ is the $i$-th elementary symmetric polynomial in the eigenvalues of an element of the Lie algebra.}
on the Lie algebra $\mathfrak{sl}(n)$, where $\mathfrak{t}$ is the  Cartan torus and $\mathcal{W}$ the Weyl group. The space \eqref{Anfamily} is non-singular. However by making a \emph{base change}, one can obtain a singular space whose resolution blows up a subset of the roots of the central ALE fiber. Coming back to the conifold: the defining equation is obtained by the non-singular family 
 \be 
 uv=z^2 + \sigma_2 
\ee
with the base change $\sigma_2=w^2$. The small resolution of the conifold blows up the simple root of $A_1$ in the central fiber. This is called a \emph{simultaneous resolution}.
The family is now fibered over the base $\mathfrak{t}$. 

For generic $n$, making such a base change resolves all the simple roots of $A_n$ in the central fiber.
One can also make a partial simultaneous resolution in which the base change maps the base $\mathfrak{t}/\mathcal{W}$ to $\mathfrak{t}/\mathcal{W}'$, where $\mathcal{W}'\subset \mathcal{W}$. In this case, the resolution of the family blows up the roots that are left invariant by $\mathcal{W}'$, in the central fiber. The base of the fibration is now parametrized by the $n-1$ $\mathcal{W}'$ invariants, that we call~$\tilde{\sigma}_i$~($i=1,...,n-1$).

In order to construct varieties with a simple flop (i.e. the small resolution blows up a single $\mathbb{P}^1$), one chooses a Weyl subgroup $\mathcal{W}'$ that leaves only one simple root invariant.  

The M-theory threefolds are obtained from these families by making the invariant coordinates $\tilde{\sigma}_i$ depend (linearly) on the parameter $w$. 

From the IIA perspective, one starts with a stack of $n$ D6-branes in flat space, that is dual to M-theory on a $A_{n-1}$ singularity. As seen before, this gives rise to a $\mathcal{N}=1$, $d=7$ $SU(n)$ gauge theory with three adjoint scalars $\phi_i$ ($i=1,2,3$).  Now we fiber this background over the $w$-plane. We do this by switching on a non-zero $w$-dependent vev for the complex adjoint scalar $\Phi=\phi_1+i\phi_2$.
This corresponds to giving angles to the D6-branes, while in M-theory one is deforming the $A_{n-1}$ singularity. The vev $\langle\Phi\rangle\in \mathfrak{sl}(n)$ breaks the $SU(n)$ gauge group on the branes to the commutant of $\langle\Phi\rangle$. In the $\mathcal{N}=1$ $d=5$ theory, the surviving massless gauge bosons live in a background vector multiplet together with the modes of the $\phi_3$ fields. A vev for $\phi_3$, with $[\langle \phi_3 \rangle, \langle \Phi \rangle]=0$, corresponds to the resolution of the singularity of the family in M-theory, that is the resolution of the roots of the ALE fiber left invariant by $\mathcal{W}'$.
Hence, the generic field $\Phi$ corresponding to a given partial simultaneous resolution is an element of the Lie algebra that commutes with the Cartan generators dual to the resolvable roots of the central fiber.

Let us consider the Dynkin diagram of $A_{n-1}$, and say that we want a simultaneous resolution of the simple root $\alpha_\ell$ ($1\leq \ell\leq n-1$). Then $\phi_3$ must live along the Cartan
\be 
\phi_3 \propto \bm  \tfrac{1}{\ell} \mathbb{1}_\ell & 0 \\ 0 & -\tfrac{1}{n-\ell}\mathbb{1}_{n-\ell}  \eem  %\in \mathfrak{sl}(n)
\ee
and the generic $\Phi$ relative to the simultaneous resolution of $\alpha_\ell$ will be 
\be \label{HiggsOnealphaAn}
\Phi = \bm  \Phi_{\ell\times \ell} & 0 \\ 0 & \Phi_{(n-\ell)\times (n-\ell)}  \eem  \in \mathfrak{sl}(n)
\ee
 The Casimirs of such Higgs field are the $\tilde{\sigma}_i$ ($i=1,...,n-1$) $\mathcal{W}'$-invariant coordinates, where $\mathcal{W}'$ is the Weyl subgroup that leaves the simple root $\alpha_\ell$ invariant.
Each block is a \emph{reconstructible Higgs} that can be written (up to a gauge transformation) in a canonical form that depends directly on the `partial' Casimirs $\tilde{\sigma}_i$ \cite{Cecotti:2010bp}.
In particular, for a  $U(m)$ block we have \cite{Cecotti:2010bp}
\be\label{recHiggsU}
\Phi_{m\times m} =\left(\begin{array}{ccccc}
0 & 1 & 0 & \cdots & 0 \\
0 & 0 & 1 & 0 & 0\\
\vdots & 0 & \ddots & \ddots & 0  \\
0 & 0 & 0 & 0 & 1 \\
(-1)^{m-1}\hat\sigma_m & (-1)^{m-2}\hat\sigma_{m-1} & \cdots & -\hat\sigma_2& \hat\sigma_1 \\
\end{array} \right)
\ee
with $\hat\sigma_j$ ($j=1,...,m$) the Casimirs of $\Phi_{m\times m}$.
The Casimirs $\tilde\sigma_i$ ($i=1,...,n-1$) of the block-diagonal total $\Phi$ are given by the collection of $\hat\sigma_j$'s of each block (with the constraint that the trace of $\Phi$ vanishes).
The Casimirs $\sigma_i$'s can be written as functions of the $\tilde{\sigma}_i$'s.

Given the total $\Phi$, whose entries depend on the Casimirs~$\sigma_i$, the family equation~\eqref{Anfamily} %(and the threefold given by $\sigma_i=\sigma_i(w)$) 
is obtained by
\be \label{AnFamiliesFromPhi}
uv = \det \left( z \mathbb{1}_n - \langle\Phi\rangle   \right)\:.
\ee
When we make the choice $\sigma_i=\sigma_i(w)$, we obtain the equation of a threefold. It is of the form of a $\mathbb{C}^\ast$ fibration. The D6-brane locus is then given by 
\be 
 \Delta_{D6} \equiv \det \left( z \mathbb{1}_n - \langle\Phi(w)\rangle   \right) = 0 \:.
\ee

The generalization to the other ADE algebras is straightforward. Given the Dynkin diagram, a simultaneous resolution corresponds to a choice of simple roots. This selects, on one side the subspace where $\phi_3$ lives, on the other side the $\mathcal{W}'$ Weyl subgroup that gives the base of the family. The commutant of $\phi_3$ tells us what is the form of a generic $\Phi$ producing the wanted family. The M-theory threefold is obtained by choosing the $w$-dependence of the invariants $\tilde{\sigma}_i$'s. This gives the map from threefolds that are 1-parameter families of deformed ADE singularities and setups of D6-branes. As for the conifold example, we will use this map to compute the HB of M-theory on such threefolds, by making computations in the dual IIA setup.

\section{Reid's pagoda}\label{Sec:Reid}
Reid's \emph{pagoda} is a class of singular CY threefolds that admit simple flops, meaning that only one exceptional $\pp^1$ is produced. It is defined as the following hypersurface:
\be\label{Eq:pagodas}
u v = z^{2 k}-w^2 \qquad \subset \quad \cc^4 \langle u, v, z, w \rangle\,.
\ee
This geometry admits $k$ normalizable deformations, and hence we expect the Higgs branch to be quaternionic $k$-dimensional. However, it is very difficult to explicitly construct the Higgs branch starting from this 3-fold data. We know from the supergravity analysis, that it is given by a $(\cc^*)^k$-fibration over $\cc^k$, whereby the base corresponds to complex structure moduli deformations, and the fiber corresponds to periods of $C_3$ on the various 3-cycles created by the deformations. But getting the global data of this variety is not straightforward.

Let us apply what seen in Section~\ref{Sec:ADEfamilies}. The threefold \eqref{Eq:pagodas} is  an ALE family  that is singular at the origin, where the ALE fiber develops an $A_{2k-1}$ singularity. Resolving the singularity blows up one exceptional $\mathbb{P}^1$, i.e. we have a simple flop. The root of the singular central fiber that is simultaneously resolved is\footnote{In the convention that $\alpha_1$ is the left root in the Dynkin diagram of $A_{2k-1}$, $\alpha_2$ is the next one and so on.} $\alpha_k$.
This M-theory background is reduced in IIA to a system of $2k$ D6-branes Higgsed by a $\Phi$ of the form
\eqref{HiggsOnealphaAn} with $n=2k$ and $\ell=k$. The $\mathcal{W}'$-invariant Casimirs are $\tilde{\sigma}_i=\hat{\sigma}_i^{(1)}$ and $\tilde{\sigma}_{i+k}=\hat{\sigma}_i^{(2)}$ with $i=1,...,k$  (and $\hat{\sigma}_1^{(1)}+\hat{\sigma}_1^{(2)}=0$).
Moreover, to obtain \eqref{Eq:pagodas} one makes the choice 
$$
\hat\sigma_j^{(1)} = \hat\sigma_j^{(2)} = 0 \,\,\,\mbox{for}\,\,\, j=1,...,k-1\qquad\mbox{and}\qquad \hat\sigma_k^{(1)} = - \hat\sigma_k^{(2)} = w
$$
Hence 
\be 
\langle \Phi \rangle = \bm J_+ & 0 \\ 0 & J_{-} \eem\,, \qquad \mbox{where}\qquad
J_{\pm }:= \bm 0& 1  &  \\ && \ddots \\ & &&&1 \\ \pm w &  & \ldots&& 0 \eem_{k} \:.
\ee
$J_{\pm }$ is the $k \times k$ Jordan block plus  $(\pm w)$ in the $(k, 1)$-entry.

The D6-branes live on a divisor in $\cc^2$ given by:
\be
\det(z \cdot \mathbb{1}_{2k}-\langle \Phi \rangle) \equiv z^{2 k}-w^2=0\,.
\ee
Notice, that this is again reducible, like for the conifold. However, the two components do not intersect transversely, so the open string spectrum will not be obvious. 
%Our approach will be to start with a stack of $2k$ coincident D6-branes at $z=0$, and switch on a particular kind of background Higgs:
%\begin{align}
%\langle \Phi \rangle &= 0 &\longrightarrow \quad \langle \Phi \rangle &= \bm J_+ & 0 \\ 0 & J_{-} \eem\\
%uv &= \det(z \cdot \mathbb{1}_{2k}-\langle \Phi \rangle) = z^{2k}& \longrightarrow \quad uv &= \det(z \cdot \mathbb{1}_{2k}-\langle \Phi \rangle) = z^{2k}-w^2
%\end{align}
%where 
%\be
%J_{\pm }:= \bm 0& 1  &  \\ && \ddots \\ & &&&1 \\ \pm w &  & \ldots&& 0 \eem_{k}
%\ee
%is the $k \times k$ Jordan block plus  $(\pm w)$ in the $(k, 1)$-entry.  

The first thing we want to study is the effective gauge group in five dimensions. We will see that it is a discrete group. Our background vev breaks the original seven-dimensional worldvolume $SU(2k)$ gauge symmetry to a subgroup given by the following $2k\times 2k$ matrices 
\be
\bm e^{i\alpha_1}\,\mathbb{1}_k &0 \\ 0 & e^{i \alpha_2}\, \mathbb{1}_k \eem \in U(1)_1 \times U(1)_2  \qquad \mbox{with} \qquad      
 \alpha_1+\alpha_2 = \frac{2\pi n}{k} \qquad n=0,1,...,k-1
 \,.
\ee
Therefore, our background Higgses
\be \label{discretegauging}
SU(2 k) \longrightarrow U(1) \times \zz_k\,.
\ee
It is generated by a continuous $U(1)$ subgroup ($n=0$) and a $\mathbb{Z}_k$ subgroup, that we can take to be
\be
\mathbb{Z}_k\: : \: \bm  e^{2 \pi i n/k} \cdot \mathbb{1}_k & \\   & \mathbb{1}_k \eem\,, \quad n = 0, \ldots, k-1\,.
\ee
So far, we have discussed the seven-dimensional perspective, and this product is the gauge group. In order to deduce the five-dimensional flavor and gauge symmetries, we proceed in two steps: first we compactify all directions transverse to the matter locus on a torus. This yields a 5d theory with \emph{gauge group} $U(1) \times \zz_k$. Now we take a decompactification limit. This will ungauge the continuous $U(1)$ factor as the volume tends to infinity. The discrete part, however, having no gauge coupling, will remain gauged from the 5d perspective. To summarize:
\be
U(1)_{\rm gauge} \times (\zz_k)_{\rm gauge} \longrightarrow U(1)_{\rm flavor} \times (\zz_k)_{\rm gauge}\,.
\ee

Now we wish to understand the Higgs branch. This consists in all possible deformations of the background Higgs field: $\Phi = \langle \Phi \rangle + \varphi$, modulo linearized gauge transformations
\be
\varphi \sim \varphi + [\langle \Phi \rangle, g]\,,
\ee
for any broken generator $g \in \mathfrak{sl}(2k)$. Let us write $g$ and $\varphi$ in the block form
\be
g= \bm \alpha & \beta \\ \gamma & \delta \eem \:, \qquad 
\varphi= \bm \varphi^\alpha & \varphi^\beta \\ \varphi^\gamma & \varphi^\delta \eem
\ee
where each block is a $k\times k$ matrix and Tr$\alpha$ = Tr$\delta=0$. Then
\be
 [\langle \Phi \rangle, g]=  \bm [J_+,\alpha] & J_+ \beta - \beta J_- \\ J_- \gamma - \gamma J_+  & [J_-,\delta] \eem\:.
\ee
We see that, due to the block-diagonal form of the Higgs vev, $\alpha$ only affects the $\varphi^\alpha$ block of $\varphi$, $\beta$ only $\varphi^\beta$, etc. This means that, in the computation of the deformations, we can work out each block individually.

Let us do it explicitly for $k=1$. We have
\be 
\varphi^\alpha\sim \varphi^\alpha + \bm \alpha_{21}-\alpha_{12} w & -2\alpha_{11}\\ 2\alpha_{11} w & -   \alpha_{21}+\alpha_{12} w \eem\:.
\ee
We can use $\alpha_{11},\alpha_{21}$ to fix the first line to zero:
\be 
\varphi^\alpha\sim \bm 0 & 0 \\  \varphi^\alpha_{21} & \varphi^\alpha_{22} \eem + \bm 0 & 0\\ \varphi^\alpha_{12} w & \varphi^\alpha_{11} \eem\:.
\ee
We see that we do not have further freedom to localize the second line, hence we do not obtain localized modes from this block. The same is true for the block related to $\delta$.
Instead, localized modes come from the off-diagonal blocks. Let us consider the block $\varphi^\beta$ and let us define its entries in the following convenient way
\be 
\varphi^\beta = \bm \varphi^\beta_L+\varphi^\beta_R & \varphi^\beta_{12} \\ \varphi^\beta_{21} & -\varphi^\beta_L+\varphi^\beta_R \eem \:.
\ee
Let us see how much we can gauge fix

\be 
\varphi^\beta\sim  \bm \varphi^\beta_L+\varphi^\beta_R & \varphi^\beta_{12} \\ \varphi^\beta_{21} & -\varphi^\beta_L+\varphi^\beta_R \eem + \bm \beta_{21}+\beta_{12} w & \beta_{22} - \beta_{11}\\ (\beta_{22} + \beta_{11}) w & -   \beta_{21}+\beta_{12} w \eem\:.
\ee
We see that we can fix to zero $\varphi^\beta_{12}$ and $\varphi^\beta_L$ by respectively choosing $(\beta_{22}-\beta_{11})$ and $\beta_{21}$, obtaining
\be 
\varphi^\beta\sim  \bm \varphi^\beta_R & 0 \\ \varphi^\beta_{21} & \varphi^\beta_R \eem + \bm \beta_{12} w & 0 \\ (\beta_{22} + \beta_{11}) w &\beta_{12} w \eem\:.
\ee
We then obtain the two modes   localized on the ideal $(w)$, i.e.
\be 
 \varphi^\beta_R,  \varphi^\beta_{21} \in \cc[w]/(w) \cong \cc \:.
\ee
These modes have charge $+1$ with respect to the $U(1)$ flavor symmetry.
Analogously we obtain two modes with $U(1)$ charge $-1$ from the block $\varphi^\gamma$.\footnote{
As done for the conifold, one can switch on a vev for the localized modes. The deformed threefold is then $uv=\det\left(z\mathbb{1}_4 + \langle\Phi\rangle + \varphi\right)$.
This provides the projection map from the Higgs branch  to the deformation space of the Pagoda with $k=2$.
}

For generic $k$ we have the same pattern. After gauge fixing we are left with $k$ constant modes in the charge $+1$ block $\varphi^\beta$ 
\be 
\varphi^\beta \sim \bm  
\varphi^\beta_k  &  0 & 0  & 0 & 0 &0  \\
 \vdots &  &  &  & & \vdots   \\
\varphi^\beta_3 &  \ldots & \ldots & \varphi^\beta_k & 0 & 0   \\
\varphi^\beta_2 & \varphi^\beta_3 & \ldots & \ldots & \varphi^\beta_k & 0   \\
\varphi^\beta_1 & \varphi^\beta_2 &\varphi^\beta_3 & \ldots & \ldots & \varphi^\beta_k 
\eem
\ee
with entries $\varphi^\beta_j \in \cc[w]/(w) \cong \cc$ ($j=1,...,k$).
Analogously, we get $k$ constant modes in the charge $-1$ block $\varphi^\gamma$.
This gives a total of  \textbf{k  hypermultiplets}.

One can also follow a different path to get the same result.
We will shift our paradigm, by relying on a physical argument put forward in \cite{Cecotti:2010bp}. First, note that the blocks $J_+$ and $J_-$ are what are called \emph{reconstructible Higgs backgrounds} in that paper. This means that their characteristic polynomials describe smooth hypersurfaces:
\be
P_{J_\pm}(z):= \det(z \cdot \mathbb{1}_{k}-J_\pm) = z^{k}\mp w\,.
\ee
Physically, each block is describing a single smooth recombined brane, with a $U(1)$ gauge group on it, let's call them $B_\pm$, sitting on the hypersurfaces:
\be
B_\pm: \quad w\mp z^k=0\,.
\ee
Now, we can turn things around, and regard this as a Higgsed background for a starting $SU(2)$ system as follows:
\be
\langle \tilde \Phi \rangle = \bm z^k & 0 \\ 0 & -z^k \eem\,.
\ee
Here, we have reinterpreted this as an initial $U(2)$ stack on $w=0$, Higgsed to $\tilde \Phi$. %This is expected to be true on the physical grounds that the smooth recombined branes must each be given by the uniquely defined $U(1)$ 7d super-Maxwell theory.

Once we have this background, we can study the fluctuations as we did before, by modding out by complexified gauge transformations. It is similar to the conifold analysis, with one important major difference, as we will see. Let us again parametrize fluctuations and gauge parameters as follows:
\be
\varphi= \bm \varphi_0 & \varphi_+\\ \varphi_- & -\varphi_0 \eem\,, \qquad g = \tfrac{1}{2}\,\bm g_0& g_+\\g_-& -g_0 \eem\,.
\ee
Now, we see that fluctuations are defined up to
\be
\varphi \sim \varphi + z^k \bm 0&g_+\\-g_-&0 \eem\,.
\ee
From this, we deduce two things: Just as for the conifold, localized modes live only in the off-diagonal part of $\varphi$. This is as expected, since we are looking at a pair of intersecting branes, albeit with a non-transversal intersection. Second and most importantly, there are now several bifundamental open string modes. Na\"ively, the modes live in the following ring:
\be
\varphi_\pm \in \cc[z]/(z^k) \cong \cc^{k}\,.
\ee
So the $(\varphi_+, \varphi_-)$ pairs can give rise to $k$ distinct hypermultiplets. This makes intuitive sense, since the two branes intersect at a \emph{fat point} of multiplicity $k$: in terms of ideals, we have
\be
(w+z^k, w-z^k) = (w, z^k)\,.
\ee
What is remarkable about this background is that the M-theory geometry sees only one vanishing $\pp^1$. Nevertheless, there are $k$ distinct membrane states that give rise to separate hypers. This point is emphasized in our companion paper \cite{Collinucci:2021wty}, where the Gopakumar-Vafa invariants corroborate this statement.

In order to really know the structure of the Higgs branch, however, we have to take into account the \emph{discrete gauge group} we found in \eqref{discretegauging}. More specifically, let us see how the $\zz_k$ discrete gauge group acts on our zero-modes:
\be
%\varphi \mapsto g^{-1} \varphi g =  \bm 0 & e^{2 \pi n i/k} \varphi_+\\ e^{-2 \pi n i/k}\varphi_- & -0 \eem\,,
\mathbb{Z}_k\::  \bm 0 &  \varphi^\beta\\ \varphi^\gamma & 0 \eem \,\,\,\mapsto\,\,\,
 \bm 0 & e^{2 \pi n i/k} \varphi^\beta\\ e^{-2 \pi n i/k}\varphi^\gamma & 0 \eem
\ee
We can conclude that RP$_k$ has a $k$-dimensional Higgs branch with the following orbifold geometry:
\be
\mathcal{H}({\rm RP}_k) = \mathbb{H}^k/\zz_k\,.
\ee
This generalizes the result found in \cite{Closset:2020scj}. The flavor symmetry of this theory will be the 7d gauge symmetry modulo the discrete 5d gauge group, so in this case:
\be
G_F = U(1)/\zz_k\,.
\ee
Note, that even though there are $k$ hypermultiplets, the flavor symmetry is of rank one. This implies that we can only switch on one real mass, if we are to think of real masses as background vev's in the usual way. The fact that only one real mass is available perfectly matches the fact that the corresponding M-theory threefold only admits a simple flop, as opposed to a reducible one.

\section{Flops of length two}

\subsection{Families of $D_n$-surfaces}

In this section we discuss threefolds that are one-parameter families of $D_n$-type ALE surfaces. As before, we call this parameter $w$. At the origin of the parameter space the surface develops a $D_n$ singularity, while on generic points this singularity is fully deformed. These examples were developed from the matrix factorizations viewpoint in \cite{Curto:aa} and from the NCCR viewpoint in \cite{Karmazyn:2017aa}.

In order to have a flop of length two, such family should have a point-like singularity at $w=0$. The resolution of such a singularity restricted to the central fiber should be the standard resolution of the trivalent node in the Dynkin diagram of $D_n$. 

To construct these threefolds explicitly one proceeds as explained in Section~\ref{Sec:ADEfamilies}: One starts with the complete family of $D_n$-type ALE surfaces over the space $\mathfrak{t}/\mathcal{W}$. 
This $(n+2)$-dimensional family ($n$-dimensional base plus $2$-dimensional fiber) is non-singular even though the fiber at the origin has a $D_n$-singularity. 

We now want to simultaneously resolve only the trivalent node of the Dynkin diagram of $D_n$, in order to generate a flop of length two \cite{Curto:aa}. To do this, one takes $\mathcal{W}'$ the Weyl subgroup corresponding to all the other simple roots
\cite{Katz:1992aa, Cachazo:2001gh}. These generate a $A_{n-3} \oplus A_1 \oplus A_1$ subalgebra of $D_n$ and the corresponding Weyl subgroup is $S_{n-2}\times \mathbb{Z}_2 \times \mathbb{Z}_2$. This subalgebra corresponds to the breaking of $SO(2n)$ to $U(n-2)\times SO(4)$, that would be produced by a Higgs $\phi_3$ along the Cartan generator dual to the trivalent root.
The Weyl invariant coordinates are the Casimirs $\sigma_i$ ($i=1,...,n-2$) of $U(n-2)$ and the Casimirs $\varpi_1$ and $\varpi_2$ of $SO(4)$. 

Now, to obtain a threefold with a flop of length two one just needs to take the $\mathcal{W}'$ invariants to depend (linearly) on the parameter $w$, in such a way that at $w=0$ all of them vanish, i.e., now $\sigma_i=\sigma_i(w)$ and $\varpi_{1,2}=\varpi_{1,2}(w)$.

\

The IIA setups that engineer these flops of length two involve not only D6-branes, but also O6$^-$-planes. These are defined such that a stack of $k$ D6/image-D6 pairs lie on top of the O-plane and carry an $SO(2n)$ gauge group. The adjoint scalar fields $\phi_i$ ($i=1,2,3$) are therefore anti-symmetric $2n \times 2n$ matrices. It will be more convenient, however, to work in a basis of $SO$ such that the Higgs has the following block diagonal structure:
\be\label{phiCahnBasis}
\phi_i = \bm A&B\\C&-A^t \eem \,, \qquad {\rm with} \quad B^t=-B\,, C^t=-C\,.
\ee
In this basis 
%Let us write $\Phi$ as a $2n\times 2n$ antisymmetric matrix in a block-diagonal form 
%\be
%\tilde\Phi = \left(\begin{array}{c|c}
%\begin{array}{cc} 0 & \tilde\Phi_{U(n-2)} \\ -\tilde\Phi_{U(n-2)}^t & 0 \\
%\end{array} 
%&0\\ \hline 0&\tilde\Phi_{SO(4)} 
%\end{array}\right)
%\ee
the matrices $G$ of $SO(2n)$ are such that
\be 
 G\cdot \mathcal{Q} \cdot G^t = \mathcal{Q}  \qquad \mbox{where} \qquad  \mathcal{Q}=
 \left(\begin{array}{c|c}
0 & \mathbb{1}_{n} \\ \hline  \mathbb{1}_{n} & 0 \\
\end{array}\right)\:.
% \left(\begin{array}{c|c}
%\begin{array}{cc} 0 & \mathbb{1}_{n-2} \\  \mathbb{1}_{n-2} & 0 \\
%\end{array} 
%&0\\ \hline 0&
%\begin{array}{cc} 0 & \mathbb{1}_{2} \\  \mathbb{1}_{2} & 0 \\
%\end{array} 
%\end{array}\right)
\ee 

The M-theory threefold is an ALE fibration, where the fibration is generated by $w$-dependent deformations of the fiber. The (simultaneous) resolution is given by a constant vev for $\phi_3$ along the Cartan dual to the (simultaneously) resolved roots. Since %\footnote{The resolution parameter sits in a 5d $\mathcal{N}=1$ (background) vector multiplet. Giving a BPS vev to the field $\Phi$ on a stack of parallel D6-branes leaves a vector multiplet in the adjont of the subgroup preserved by $\Phi$. The reduction of $\phi_3$ to 5d sits in such a vector multiplet.}
$[\Phi,\phi_3] = 0$,
$\Phi$ should live in the adjoint representation of $U(n-2)\times SO(4)\subset SO(2n)$. 
The Higgs field is then the following $2n\times 2n$ matrix (in the basis \eqref{phiCahnBasis}) 
\be
\Phi = \left(\begin{array}{cc|cc}
\Phi_{U(n-2)} &  &  & \\   & a  &  &  b \\ \hline   & & -\Phi_{U(n-2)}^t & \\   & c &  & -a^t \\
\end{array} \right)
\ee
where the block $\Phi_{SO(4)}=\bm a & b \\ c & - a^t  \eem$, with $b,c$ antisymmetric $2\times 2$ matrices, is a field in the adjoint of $SO(4)$, while $\Phi_{U(n-2)}$ is in the adjoint of $U(n-2)$.

The fields $\Phi_{SO(4)}$ and $\Phi_{U(n-2)}$ depend on the Casimirs of the corresponding groups; in particular $\varpi_{2}=\sqrt{\det(\Phi_{SO(4)})}$ and $\varpi_1=\frac12 $Tr$\Phi_{SO(4)}^2 + 2 \varpi_2$.  There is an analogous relation to \eqref{AnFamiliesFromPhi} that gives the M-theory threefold in terms of the IIA Higgs field \cite{Cachazo:2001gh}:
\be\label{D2ndefFamily}
x^2 + z y^2 - \frac{\sqrt{\det(z\mathbb{1} + \Phi^2)}  - \varpi_2^2\sigma_{n-2}^2  }{z} + 2  \varpi_2\sigma_{n-2} y=0
\ee
This is a $\mathbb{C}^\ast$ fibration. The $\mathbb{C}^\ast$ fiber degenerates over the D6-brane locus (given by the discriminant of the quadric in $y$):
\be
 \Delta_{D6} \equiv \sqrt{\det(z\mathbb{1} + \Phi^2)} = 0
\ee
The O6-plane locus is at $z=0$ (where the coefficient of $y^2$ vanishes). The type IIA target space is a double cover of the base of the $\mathbb{C}^\ast$ fibration, that can be given by the equation $\xi^2=z$ (see \cite{Collinucci:2021wty} for more details).

For the following, it is more convenient  to exchange rows and columns to bring $\Phi$ into the form
\be\label{PhiDnSimpleForm}
\Phi = \left(\begin{array}{c|c}
\begin{array}{cc}  \Phi_{U(n-2)} &   \\   & -\Phi_{U(n-2)}^t \\
\end{array} 
&0\\ \hline &\Phi_{SO(4)} 
\end{array}\right).
\ee
In this basis,  the elements $g$ of the algebra $\mathfrak{so}(2n)$ satisfy:
\begin{equation}
g Q+Qg^t = 0, \qquad\mbox{with}	\qquad
Q = \left(
\begin{array}{cc|cc}
0 & \mathbb{1}_{n-2} &  & \\
 \mathbb{1}_{n-2} & 0 & &  \\
\hline
  &  & 0 & \mathbb{1}_{2} \\
  &  & \mathbb{1}_{2} & 0 \\
\end{array}
\right)\:.
\end{equation}

Each block can be written (by a gauge transformation) in a canonical form, where the entries directly depend on the Casimirs. In particular, for the $U(n-2)$ block we have the form \eqref{recHiggsU}.
For the $SO(4)$ block, one can use that $\mathfrak{so}(4)=\mathfrak{su}(2)\oplus \mathfrak{su}(2)$ to obtain the canonical form 
\begin{equation}\label{recHiggsSO4}
\Phi_{SO(4)}=\left(
\begin{array}{cccc}
  0 & 1 & 0 & -\frac{\varpi_1}{4} \\
  \frac{\varpi_1}{4}- \varpi_2 & 0 & \frac{\varpi_1}{4} & 0  \\
  0 & 1 & 0 & \varpi_2-\frac{\varpi_1}{4} \\
  -1 & 0 & -1 & 0  \\
\end{array}
\right),
\end{equation}
where the Casimirs of the two $SU(2)$ are $\frac{\varpi_1}{4}$ and $\varpi_2-\frac{\varpi_1}{4}$.

\subsection{Brown-Wemyss threefold}\label{Sec:Brown-Wemyss}

In this section we study a one-parameter family of deformed $D_4$ singularities, i.e. $n=4$. Moreover we will set $\varpi_2=0$. When this happens the family takes the simple form
\be\label{D4defFamily}
 x^2 + z y^2 - \left( z	\sigma_1^2 + (z-\sigma_2)^2  \right)\left( z+\varpi_1 \right) = 0\:.
\ee

The threefold is defined by the following $w$-dependence of the Casimirs of $U(2)\times SO(4)$:
\begin{equation}
\sigma_1 = -w\,, \qquad \sigma_2 = w\,, \qquad \varpi_1= - w \,, \qquad \varpi_2=0\:.
\end{equation} 
Plugging these into \eqref{D4defFamily}, one obtains the hypersurface
\begin{equation}\label{Eq:Brown Wemyss}
x^2 + z y^2 - \left( z-w \right) \left( z	w^2 + (z-w)^2  \right) = 0  \:.
\end{equation}
This threefold was introduced in \cite{BrownWemyss}. It is singular at the origin, where the ALE fiber develops a $D_4$ singularity. The total space admits a small resolution with a flop of length two.
This threefold has Milnor number
%\footnote{As a zero-dimensional subscheme of $\mathbb C^2 \ni (w,z)$, the Jacobi algebra is supported both on the origin, and on the point $(w = 27/8, z = - 27/16)$, and its length is  12. Considering only the component supported on the origin, we get 11 as Milnor number.} \todo{serve la footnote?} 
11 and the number of normalizable deformations is $6$, hence we expect a 6-dimensional Higgs Branch. Moreover, the threefold admits a small resolution, leading to a $U(1)$ flavor symmetry.\footnote{See \cite{Collinucci:2018aho,Collinucci:2019fnh} for ax explicit resolution of these geometries by quiver techniques with a focus on the $U(1)$ symmetry and its charges}

In IIA we start with an $SO(8)$ stack at the orientifold location $z=0$ and we switch on a  
background Higgs 
\begin{equation}\label{EqPhiBW}
\Phi=\left(
\begin{array}{cccc|cccc}
 0 & 1 & 0 & 0 & 0 & 0 & 0 & 0 \\
 -w & -w & 0 & 0 & 0 & 0 & 0 & 0 \\
% \hline
 0 & 0 & 0 & w & 0 & 0 & 0 & 0 \\
 0 & 0 & -1 & w & 0 & 0 & 0 & 0 \\
 \hline
 0 & 0 & 0 & 0 & 0 & 1 & 0 & \frac{w}{4} \\
 0 & 0 & 0 & 0 & -\frac{w}{4} & 0 & -\frac{w}{4} & 0 \\
 0 & 0 & 0 & 0 & 0 & 1 & 0 & \frac{w}{4} \\
 0 & 0 & 0 & 0 & -1 & 0 & -1 & 0 \\
\end{array}
\right).
\end{equation}

The group that commutes with the Higgs field $\Phi$ is isomorphic to $U(1)\times \mathbb{Z}_2 \times \mathbb{Z}_2$:
\begin{equation}\label{u1}
\left(
\begin{array}{cccc|ccc}
 e^{i\alpha} & 0 & 0 & 0 & 0 & 0  & 0 \\
 0 & e^{i\alpha} & 0 & 0 & 0 & 0  & 0 \\
 0 & 0  & e^{-i\alpha} & 0 &  0 & 0 & 0 \\
0 & 0 & 0   & e^{-i\alpha} & 0 & 0 & 0 \\
\hline
 0 &0 & 0 &  0 & & & \\
 0 &0  & 0 &  0& & \boldsymbol{\varrho} &  \\
 0 &0  & 0 &  0& & & \\
\end{array}
\right)
\end{equation}
where  $\alpha$ is a phase and $\boldsymbol{\varrho}$ live in the $ \mathbb{Z}_2 \times \mathbb{Z}_2\subset SO(4)$ generated by
\begin{equation}\label{sigma}
\setlength{\jot}{16pt} 
\begin{split}
& \boldsymbol{\zeta}_1 = \left(\begin{array}{cccc}
0 & 0 & 1 & 0 \\
0 & 1 & 0 & 0 \\
1 & 0 & 0 & 0 \\
0 & 0 & 0 & 1 \\
\end{array} \right) \hspace{1cm}\mbox{and}  \hspace{1cm}
\boldsymbol{\zeta}_2 = \left(\begin{array}{cccc}
-1 & 0 & 0 & 0 \\
0 & -1 & 0 & 0 \\
0 & 0 & -1 & 0 \\
0 & 0 & 0 & -1 \\
\end{array} \right) \:.  \\
\end{split}
\end{equation}
We then have a continuous abelian group, that is seen as a flavor group in the 5d theory, times a discrete gauge group.

We now consider the Higgs branch. As for the examples in Section~\ref{Sec:Reid}, this consists in the deformations $\varphi$ of the background Higgs, modulo linearized $SO(8)$ gauge transformations:
\begin{equation}
\varphi \sim \varphi + [ \Phi, g ] \:.
\end{equation}
The commutator  $[ \Phi, g ]$ can be written in the block form
\begin{equation}
[\Phi,g] =  \left(
\begin{array}{ccc|ccc|ccc}
& & & & & & & & \\
& B_{2\times 2} & & &A_{2\times 2}^{u} &  & &C_{2\times 4} & \\
& & & & & & & & \\
\hline
& & & & & & & & \\
& A_{2\times 2}^{d} & & & -B_{2\times 2}^t & & & D_{2\times 4} & \\
& & & & & & & & \\
\hline
& & & & & & & & \\
& D_{4\times 2}& & & C_{4\times 2} & & & -B_{4\times 4}& \\
& & & & & & & & \\
\end{array}
\right)
\end{equation}
where $C_{2\times 4}$ is completely determined by $C_{4\times 2}$  (analogously for the $D$-blocks). Due to the block-diagonal form of the Higgs \eqref{EqPhiBW}, each block of  $[ \Phi, g ]$ depends only on the entries of $g$ in the same block.

Let us proceed now block by block. We start with
$$B_{2\times 2}=\left(
\begin{array}{cc}
 g_{21}+g_{12} w & -g_{11}+g_{22}+g_{12} w \\
 -(g_{11}+g_{21}-g_{22}) w & -g_{21}-g_{12} w \\
\end{array}
\right)\:.$$
Using $g_{21}$ and the combination $g_{11}-g_{22}$ we can fix to zero the corresponding entries $\varphi_{11}$ and $\varphi_{12}$ in the fluctuation of the Higgs.
We are then left with:
\begin{equation}
\varphi_{2\times 2} \sim \underbrace{\left(
\begin{array}{cc}
0 & 0 \\
\varphi_{21} &\varphi_{22}  \\
\end{array}
\right)}_{\varphi_{2\times 2}}+\underbrace{\left(
\begin{array}{cc}
0 & 0 \\
-w(\varphi_{12}-\varphi_{11}) & \varphi_{11} \\
\end{array}
\right)}_{B_{2\times 2}}.
\end{equation}
As a result we see that $\varphi_{21}$ and $\varphi_{22}$ are not  constrained, and so they are not dynamical in 5d.

The other relevant diagonal block
$${\scriptsize B_{4\times 4} =\left(
\begin{array}{cccc}
 g_{58}+g_{65}+\frac{1}{4} (g_{56}-g_{76}) w & g_{66}-g_{55} & 0 & -\frac{1}{4} (g_{55}+g_{66}) w \\
 \frac{1}{4} (g_{66}-g_{55}) w & g_{58}-g_{65}+\frac{1}{4} (-g_{56}-g_{76}) w & \frac{1}{4} (g_{55}+g_{66}) w & 0 \\
 0 & g_{55}+g_{66} & -g_{58}-g_{65}+\frac{1}{4} (g_{76}-g_{56}) w & \frac{1}{4} (g_{55}-g_{66}) w \\
 -g_{55}-g_{66} & 0 & g_{55}-g_{66} & -g_{58}+g_{65}+\frac{1}{4} (g_{56}+g_{76}) w \\
\end{array}
\right)}$$
does not generate localized modes as well. In fact, using the combinations $(g_{66}-g_{55})$, $(g_{66}+g_{55})$, $(g_{58}-g_{65})$ and $(g_{58}+g_{65})$ we can set to zero, for example, the entries $\varphi_{55}$, $\varphi_{56}$, $\varphi_{66}$ and $\varphi_{76}$, remaining with:
\begin{equation}
\varphi_{4\times 4} \sim \underbrace{\left(\begin{array}{cccc}
0 & 0 & 0 & \varphi_{58} \\
\varphi_{65} & 0 & -\varphi_{58} & 0 \\
0 & 0 & 0 & \varphi_{78} \\
0 & 0 & 0 & 0 \\
\end{array} \right)}_{\varphi_{4\times 4}}+\underbrace{\left(
\begin{array}{cccc}
 0& 0 & 0 & -\frac{w \varphi_{76}}{4}  \\
 \frac{w \varphi_{56}}{4} & 0 & \frac{w \varphi_{76}}{4} & 0 \\
 0 & 0 & 0 & - \frac{w \varphi_{56}}{4}\\
0 & 0 & 0 & 0 \\
\end{array}
\right)}_{B_{4\times 4}} \:.
\end{equation}

The first localized mode comes when we consider $A_{2\times 2}^u$. The gauge equivalence is
\begin{equation}
\varphi_{2\times 2}^u \sim \underbrace{\left(
\begin{array}{cc}
 0 & \varphi_{14} \\
-\varphi_{14}& 0 \\
\end{array}
\right)}_{\varphi_{2\times 2}^u}+\underbrace{\left(
\begin{array}{cc}
 0 & -w g_{14} \\
 w g_{14} & 0 \\
\end{array}
\right)}_{A_{2\times 2}^u}.
\end{equation}
We immediately see that $\varphi_{14}$ is localized on the ideal $(w)$, giving:
\begin{equation}
\varphi_{14}\in \mathbb{C}[w]/(w) \cong \mathbb{C}
\end{equation}
and so it corresponds to 1 localized 5d mode. We note that this mode has charge $+2$ with respect to the flavor $U(1)$ in (\ref{u1}). 

The block $A_{2\times 2}^d$ acts analogously to $A_{2\times 2}^u$ and it yields a localized mode with charge~$-2$ with respect to the flavor $U(1)$.

Let us come to the block $C_{4\times 2} $.
The gauge equivalence is:
\begin{equation}
\varphi_{4\times 2} \sim \underbrace{ \left(
\begin{array}{cc}
 \varphi_{53} & \varphi_{54} \\
 \varphi_{63} & \varphi_{64} \\
 \varphi_{73} & \varphi_{74} \\
 \varphi_{83} & \varphi_{84} \\
\end{array}
\right)}_{\varphi_{4\times 2}} +\underbrace{ \left(
\begin{array}{cc}
 -g_{18}-g_{27}-\frac{g_{16} w}{4} & \left(g_{17}+g_{27}-\frac{g_{26}}{4}\right) w-g_{28} \\
 \frac{1}{4} (g_{15}+g_{17}) w-g_{28} & \frac{1}{4} (4 g_{18}+g_{25}+g_{27}+4 g_{28}) w \\
 -g_{18}-g_{25}-\frac{g_{16} w}{4} & \left(g_{15}+g_{25}-\frac{g_{26}}{4}\right) w-g_{28} \\
 g_{15}+g_{17}-g_{26} & g_{25}+g_{27}+(g_{16}+g_{26}) w \\
\end{array}
\right)}_{C_{4\times 2}}.
\end{equation}
Almost all the entries in $\varphi$ corresponding to this block can be gauge-fixed to zero, except $\varphi_{64}$ and two linear combinations of $\varphi_{54}$, $\varphi_{63}$, $\varphi_{64}$.
After having fixed all the other entries to zero, we have:
\begin{equation}
\begin{split}
& \varphi_{54} \sim \varphi_{54} +w^2 \left(-\frac{g_{16}}{2}-\frac{g_{26}}{2}\right)+w \left(g_{17}-\frac{g_{26}}{4}+\frac{\varphi_{53}}{2}-\frac{\varphi_{73}}{2}-\frac{\varphi_{84}}{2}\right)-g_{28} \\
& \varphi_{63} \sim \varphi_{63} +\frac{1}{4} w (g_{26}-\varphi_{83})-g_{28} \\
& \varphi_{64} \sim \varphi_{64} +\frac{g_{26} w^2}{4}+\frac{1}{4} w (4 g_{28}+2 \varphi_{53}+2 \varphi_{73}+\varphi_{84}) \\
& \varphi_{74} \sim \varphi_{74} +w^2 \left(-\frac{g_{16}}{2}-\frac{g_{26}}{2}\right)+w \left(-g_{17}+\frac{3 g_{26}}{4}-\frac{\varphi_{53}}{2}+\frac{\varphi_{73}}{2}-\varphi_{83}-\frac{\varphi_{84}}{2}\right)-g_{28} \\
\end{split}\nonumber
\end{equation}
To make the computation easier and without loss of generality, we redefine $\varphi_{54}$, $\varphi_{63}$ and $\varphi_{74}$ as 
\begin{equation}
\varphi_{54} = \psi_1-\psi_2 \,,\qquad \varphi_{63} = \psi_3 \,,\qquad  \varphi_{74} = \psi_1+\psi_2 \:.
\end{equation}
Using the gauge freedom given by $g_{28}$ we can set $\psi_3$ to zero, remaining with:\footnote{And discarding the dependence on the other $\varphi_{ij}$, that are not free parameters}
\begin{equation}
\begin{split}
& \varphi_{64} \sim \varphi_{64} +\frac{g_{26} w^2}{2} \\
& \psi_1 \sim \psi_1+ \frac{1}{4} w^2 (-2 g_{16}-2 g_{26}) \\
& \psi_2 \sim \psi_2 +\frac{1}{4} w (2 g_{26}-4 g_{17}) \\
\end{split}
\end{equation}
We immediately see that $\varphi_{64}$ is localized on the ideal $(w^2)$, yielding:
\begin{equation}
\varphi_{64}\in \mathbb{C}[w]/(w^2) \cong \mathbb{C}^2 \:.
\end{equation}
On the other hand, we see that:
\begin{equation}
\begin{split}
& \psi_1\in \mathbb{C}[w]/(w^2) \cong \mathbb{C}^2 \\
& \psi_2\in \mathbb{C}[w]/(w) \cong \mathbb{C} \\
\end{split}
\end{equation}
We then have a total of 5 localized modes with charge $+1$ under the $U(1)$ in (\ref{u1}).

The block $D_{4\times 2}$ works like $C_{4\times 2}$ and gives 5 localized modes with charge~$-1$ with respect to the $U(1)$ in (\ref{u1}).

\indent Summing up, we obtain a 5d $\mathcal{N}=1$  theory with six hypermultiplets  (the modes with opposite charge pair up into a hyper):
\begin{itemize}
\item \textbf{1 hyper} of charge 2:\footnote{The existence of a charge-2 state localized at the origin of threefolds admitting flops of length two was already predicted in \cite{Collinucci:2018aho}.}
$$
\bm Q_0 \\ \tilde Q_0 \eem = \bm \varphi_{14} \\ \varphi_{41} \eem
$$
\item \textbf{5 hypers} of charge 1:
$$
\bm Q_1 \\ \tilde Q_1 \eem = \bm \varphi_{64}^{(1)} \\ \varphi_{28}^{(1)} \eem,\qquad
\bm Q_2 \\ \tilde Q_2\eem = \bm \varphi_{64}^{(2)} \\ \varphi_{28}^{(2)} \eem,\qquad
$$
$$
\bm Q_3 \\ \tilde Q_3 \eem = \bm \psi_{1}^{(1)} \\ \tilde\psi_{1}^{(1)} \eem,\qquad
\bm Q_4 \\ \tilde Q_4 \eem = \bm \psi_{1}^{(2)} \\ \tilde\psi_{1}^{(2)} \eem,\qquad
\bm Q_5 \\ \tilde Q_5 \eem = \bm \psi_{2} \\ \tilde\psi_{2} \eem.
$$
\end{itemize}

The Higgs branch will then be $\mathbb{H}^6$ modded out by the discrete gauge symmetry $\mathbb{Z}_2 \times \mathbb{Z}_2$. 

Let us analyze the action of  the generators $\boldsymbol{\zeta}_1$ and $\boldsymbol{\zeta}_2$ of $\mathbb{Z}_2 \times \mathbb{Z}_2$ on the zero modes we have just  found.
\begin{itemize}
\item The charge-2 hyper $(Q_0,\tilde Q_0)$ is unaffected by the  $\mathbb{Z}_2 \times \mathbb{Z}_2$.
\item The non-trivial action occurs on the charge-1 hypers. The gauge fixed $\varphi_{4\times 2}$  block~is 
\begin{equation}
\varphi_{4\times 2} = \left(
\begin{array}{cc}
0 & \psi_1- \psi_2 \\
0& \varphi_{64} \\
0 & \psi_1+\psi_2 \\
0 & 0 \\
\end{array}
\right).
\end{equation}
The generator $\boldsymbol{\zeta}_2$ changes sign to all the modes, thus 
$$
\boldsymbol{\zeta}_2 \,: \bm Q_i\\ \tilde Q_i\eem  \mapsto - \bm Q_i\\ \tilde Q_i\eem \qquad i=1,...,5 \:.
$$
On the other hand, $\boldsymbol{\zeta}_1$ swaps the first and the third row, thus giving:
$
 \varphi_{64} \mapsto \varphi_{64}$, $
\psi_1 \mapsto \psi_1$, $
\psi_2 \mapsto -\psi_2$, i.e.
$$
\boldsymbol{\zeta}_1 \,: \bm Q_i\\ \tilde Q_i\eem  \mapsto  \bm Q_i\\ \tilde Q_i\eem \qquad i=1,...,4 \qquad\mbox{and}\qquad \bm Q_5\\ \tilde Q_5\eem  \mapsto - \bm Q_5\\ \tilde Q_5\eem.
$$
\end{itemize}

We finally claim that M-theory on the threefold \eqref{Eq:Brown Wemyss} leads to a 5d theory with Higgs branch
\begin{equation}
\mathcal{M}_{\rm HB} = \mathbb{H} \times \left(  \mathbb{H}^4\times \mathbb{H}/\mathbb{Z}_2      \right)/\mathbb{Z}_2 = \mathbb{H} \times \mathbb{H}^4/\mathbb{Z}_2\times \mathbb{H}/\mathbb{Z}_2     
\end{equation}
where the $\mathbb{Z}_2$ inverts the coordinates.
The flavor symmetry is the 7d gauge symmetry,~i.e. 
\be 
 G_F = U(1)\:.
\ee
In this case, we do not have to mod the $U(1)$ by the discrete symmetry, since $\mathbb{Z}_2\times \mathbb{Z}_2$ acts differently on the $U(1)$-charged zero modes (there is no element of $U(1)$ that acts on all the modes equally to an element of $\mathbb{Z}_2\times\mathbb{Z}_2$).

\subsection{Laufer threefold}

We now generalize the computation done in the previous section to a famous flop of length two, first discovered by Laufer \cite{laufer}. It is given by the following hypersurface:
\begin{equation}\label{laufer}
x^{2}+z y^{2}-t\left(t^{2}+z^{2 k+1}\right)=0 \quad \text { with } k \geq 1\:.
\end{equation}
By making the change of variable $t=w-z$, one can put this threefold in the form of a $D_{2k+3}$ family with\footnote{The fact that $\sigma_{2k-1} = 1$ makes the singularity of the ALE fiber at the origin be $D_{2k+2}$.}
\begin{equation}\label{sigmaivarpiLaufer}
\sigma_{2k+1}=w, 
\qquad \sigma_{2k-1} = 1, \qquad \sigma_{i\neq 2k\pm1}= 0, \qquad
\varpi_1= - w , \qquad \varpi_2=0 \:.
\end{equation} 
The form of the Higgs field is like in \eqref{PhiDnSimpleForm} with $n=2k+3$. The $SO(4)$ block is identical to the one for the Brown-Wemyss threefold (see \eqref{EqPhiBW}). The $U(2k+1)$ block is given by \eqref{recHiggsU} with the Casimirs defined in \eqref{sigmaivarpiLaufer}.

The preserved group is again the diagonal $U(1)$ of the $U(2k+1)$ block times the~$\mathbb{Z}_2\times\mathbb{Z}_2$ generated by \eqref{sigma}.

The computation of the zero modes proceeds analogous to what done in Section~\ref{Sec:Brown-Wemyss}. Now the linearized gauge variation of the deformation $\varphi$ is
\begin{equation}
[\Phi,g] =  \left(
\begin{array}{ccc|ccc|ccc}
& & & & & & & & \\
& B_{(2k+1)\times (2k+1)} & & &A_{(2k+1)\times (2k+1)}^{u} &  & &C_{ (2k+1)\times 4} & \\
& & & & & & & & \\
\hline
& & & & & & & & \\
& A_{(2k+1)\times (2k+1)}^{d} & & & -B_{(2k+1)\times (2k+1)}^t & & & D_{ (2k+1)\times 4} & \\
& & & & & & & & \\
\hline
& & & & & & & & \\
& D_{4\times (2k+1)}& & & C_{4\times (2k+1)} & & & -B_{4\times 4}& \\
& & & & & & & & \\
\end{array}
\right) \:.
\end{equation}
For the zero modes, one again checks if the various blocks of  $[\Phi,g] $ localize any mode in~5d. We find:
\begin{itemize}
\item[$\triangleright$] $B_{(2k+1)\times (2k+1)}$ and $B_{4\times 4}$ do not localize any modes.
\item[$\triangleright$] $A_{(2k+1)\times (2k+1)}^{u}$ localizes one entry $\varphi_2$ as:
\begin{equation}
\varphi_2\in \mathbb{C}[w]/(w^k) \cong \mathbb{C}^k,
\end{equation}
thus yielding $k$ charge $2$ localized modes. The same goes for $A_{(2k+1)\times (2k+1)}^{d}$, from which we obtain $k$ modes $\tilde\varphi_2$ of charge $-2$.
\item[$\triangleright$] $C_{4\times (2k+1)}$ localizes three entries with the same pattern as in the Brown-Wemyss case, namely:
\begin{equation}
\begin{split}
& \varphi_{1}\in \mathbb{C}[w]/(w^{k+1}) \cong \mathbb{C}^{k+1} \\
& \psi_1\in \mathbb{C}[w]/(w^{k+1}) \cong \mathbb{C}^{k+1} \\
& \psi_2\in \mathbb{C}[w]/(w) \cong \mathbb{C} \\
\end{split},
\end{equation}
obtaining a total of $2k+3$ charge $1$ localized modes. $D_{4\times (2k+1)}$ gives the same matter content as $C_{4\times (2k+1)}$, but with charge $-1$.
\end{itemize}

Summarizing, the spectrum is given as follows:
\begin{itemize}
\item \textbf{$\boldsymbol{k}$ hypers} of charge 2:
$$
\bm Q_i \\ \tilde Q_i \eem = \bm \varphi_{2}^{(i)} \\ \tilde\varphi_{2}^{(i)} \eem \qquad i=1,...,k\:;
$$
\item \textbf{$\boldsymbol{2k+3}$ hypers} of charge 1:
$$
\bm Q_{i+k} \\ \tilde Q_{i+k} \eem = \bm \varphi_{1}^{(i)} \\ \tilde\varphi_{1}^{(i)}  \eem,
\,\, \bm Q_{i+2k+1} \\ \tilde Q_{i+2k+1} \eem = \bm \psi_{1}^{(i)} \\ \tilde\psi_{1}^{(i)}  \eem
\,\, i=1,...,k+1\:, \,\mbox{ and }\, \bm Q_{3k+3} \\ \tilde Q_{3k+3} \eem = \bm \psi_{2} \\ \tilde\psi_{2}\eem \:.
$$
\end{itemize}

Like for the Brown-Wemyss' case the discrete group in (\ref{sigma}) acts in the following way:
\begin{itemize}
\item The charge 2 hypers are unaffected by the  $\mathbb{Z}_2 \times \mathbb{Z}_2$ discrete symmetry.
\item The charge 1 hypers transform under $\boldsymbol{\zeta}_2$ as
$$
\boldsymbol{\zeta}_2 \,: \bm Q_i\\ \tilde Q_i\eem  \mapsto - \bm Q_i\\ \tilde Q_i\eem \qquad i=k+1,...,3k+3 \:.
$$
The generator $\boldsymbol{\zeta}_1$ acts as
$$
\boldsymbol{\zeta}_1 \,: \bm Q_i\\ \tilde Q_i\eem  \mapsto  \bm Q_i\\ \tilde Q_i\eem \qquad i=k+1,...,3k+2 \qquad\mbox{and}\qquad \bm Q_{3k+3}\\ \tilde Q_{3k+3}\eem  \mapsto - \bm Q_{3k+3}\\ \tilde Q_{3k+3}\eem.
$$
\end{itemize}
The Higgs branch of M-theory on Laufer's threefold is then
\begin{equation}
\mathcal{M}_{\rm HB} = \mathbb{H}^k \times \left(  \mathbb{H}^{2k+2}\times \mathbb{H}/\mathbb{Z}_2      \right)/\mathbb{Z}_2 = \mathbb{H}^k \times \mathbb{H}^{2k+2}/\mathbb{Z}_2\times \mathbb{H}/\mathbb{Z}_2     
\end{equation}
where the $\mathbb{Z}_2$ inverts the coordinates.
Like in the previous example, the flavor group is $G_F=U(1)$.

\section{Non-simple flops}

In this section we present an easy example of non-simple flop, i.e. an isolated singularity whose exceptional locus is a collection of $\mathbb{P}^1$'s. This threefold is given by the equation
\begin{equation}\label{generalized conifold}
uv=z^3-w^2z
\end{equation}
and was dubbed `generalized conifold' in \cite{Aspinwall:2010mw}.
Since it is a $A_2$-family, the corresponding IIA Higgs field lives in the adjoint of $SU(3)$. It is given by
\begin{equation}
\Phi = \left(
\begin{array}{ccc}
 0 & 0 & 0 \\
 0 & -w & 0 \\
 0 & 0 & w \\
\end{array}
\right).
\end{equation}
As for the Reid's pagodas (that were also $A_n$-families) we recover the hypersurface (\ref{generalized conifold}) computing:
\begin{equation}
uv = \text{det}(z\mathbb{1}-\Phi) = z^3-w^2z \:.
\end{equation}
The group preserving the above Higgs is:
\begin{equation}\label{gauge}
U=\left(
\begin{array}{ccc}
 e^{i\alpha} & 0 & 0 \\
 0 & e^{i\beta}  & 0 \\
 0 & 0 &  e^{i\gamma}  \\
\end{array}
\right) \hspace{1cm} \text{with } \alpha+\beta+\gamma = 0 \,\, {\rm mod}\,\, 2\pi\:,
\end{equation}
where we have decoupled the diagonal center of mass $U(1)$.\\
\indent In order to find the possible zero modes we must mod the fluctuations $\varphi$ of the Higgs by:
\begin{equation}
\varphi \sim \varphi + [\Phi,g]
\end{equation}
with $g$ a generic element of $\mathfrak{sl}(3)$.

A direct computation shows that:
\begin{equation}
\varphi \sim \underbrace{\left(
\begin{array}{ccc}
\varphi_{11} &  \varphi_{12} &  \varphi_{13} \\
\varphi_{21} &  \varphi_{22} &  \varphi_{23} \\
\varphi_{31} &  \varphi_{32} &  \varphi_{33} \\
\end{array}
\right)}_{\varphi}+\underbrace{\left(
\begin{array}{ccc}
 0 & g_{12} w & -g_{13} w \\
 -g_{21} w & 0 & -2 g_{23} w \\
 g_{31} w & 2 g_{32} w & 0 \\
\end{array}
\right)}_{[\Phi,g]} \:.
\end{equation}
We then see that $\varphi_{11}$, $\varphi_{22}$ and $\varphi_{33}$ cannot be fixed, and so do not give rise to localized modes in 5d.
On the other hand, we note that using the gauge freedom we can set:
\begin{equation}
\varphi_{12}, \varphi_{13}, \varphi_{23} \in \mathbb{C}[w]/(w) \cong \mathbb{C}
\end{equation}
so that we obtain 3 localized modes. The same goes for $\varphi_{21}, \varphi_{31}$ and $\varphi_{32}$, that give rise to other 3 modes.

It is immediate to obtain the charges of the modes under the three dependent $U(1)$s in (\ref{gauge}):
\begin{equation}
U\varphi U^{-1} = \left(
\begin{array}{ccc}
 0 & e^{i (\alpha -\beta )} \varphi_{12} & e^{i (\alpha -\gamma )} \varphi_{13} \\
 e^{-i (\alpha -\beta )} \varphi_{21} & 0 & e^{i (\beta -\gamma )} \varphi_{23} \\
 e^{-i (\alpha -\gamma )} \varphi_{31} & e^{-i (\beta -\gamma )} \varphi_{32} & 0 \\
\end{array}
\right)\:.
\end{equation}
There is no discrete gauge symmetry and then the Higgs branch is simply given by \textbf{three free hypermultiplets}. Hence, the Higgs branch is simply:
\be
\mathcal{M}_{\rm HB}  = \mathbb{H}^3\,,
\ee
with flavor symmetry
\be
G_F = U(1)^2\,.
\ee

\section{Non-resolvable threefolds}

In this last section we study a couple of simple examples of threefolds that have terminal singularities, i.e. singularities that cannot be crepantly resolved. 

\subsection{T-brane data}
We start with the class of threefolds given by the equation
\be 
 uv = z^{2k+1} + w^2
\ee
These are $A_{2k}$-families and then admit a description in IIA in terms of a non-zero vev for a Higgs field living on a $SU(2k+1)$ stack of D6-branes.

Let us describe in more detail the simplest case, i.e. $k=1$:
\be \label{non-resol-simpleone}
 uv = z^{3} + w^2 \:.
\ee
This manifold was studied in \cite{Braun:2014nva} in the context of F-theory, where the authors showed that there is matter localized at the singularity, even though such isolated singularity does not admit a crepant small resolution. It admits however a non-K\"ahler resolution, as anticipated in \cite{Grimm:2011tb}.

Here we confirm the existence of one localized hyper. The characteristic polynomial is now singular, hence the field $\Phi$ does not take the form \eqref{recHiggsU} of a reconstructible Higgs. However we can work out the form of $\Phi$ that deforms the $SU(3)$ stack to \eqref{non-resol-simpleone}:
\begin{equation}
\Phi = \left(
\begin{array}{ccc}
 0 & w & 0 \\
 0 & 0 & w \\
 1 & 0 & 0 \\
\end{array}
\right) \:.
\end{equation}
In order to find the zero-modes in the fluctuation matrix $\varphi$ we have to mod out by gauge equivalences:
\begin{equation}
\varphi \sim \varphi +[\Phi,g]
\end{equation}
where $g \in \mathfrak{gl}(3)$.

Doing so, we find that all the entries in $\varphi$ can be fixed to zero or are not localized on any ideal, except for:
\begin{equation}
\begin{split}
& \varphi_{12}\sim \varphi_{12} +w(g_{22}-g_{33}) \hspace{1cm}  \varphi_{23}\sim \varphi_{23}- w(g_{22}-g_{33})  \\
& \varphi_{22}\sim \varphi_{22}+ w(g_{32}-g_{21}) \hspace{1cm} \varphi_{33}\sim \varphi_{33}- w(g_{32}-g_{21}) \\
\end{split}
\end{equation}
We note that $\varphi_{12}$ and $\varphi_{23}$ depend on the same parameter, as $\varphi_{22}$ and $\varphi_{33}$ do. As a consequence, we can choose to localize the first two (say $\varphi_{12}$ and $\varphi_{22}$) in 5d, while the other stay non-dynamical.
Acting in this way we get:
\begin{equation}
 \varphi_{12} ,  \varphi_{22}\in \mathbb{C}[w]/(w) \cong \mathbb{C}
\end{equation}
thus giving us 1+1 = 2 modes in total. Since there is no discrete gauge symmetry left by the Higgs vev, the Higgs branch is given by \textbf{one free hypermultiplet}. 

This computation is easily generalized to a generic $k$. The Higgs field is now
\begin{equation}
\Phi = \left(
\begin{array}{ccc}
  & w & 0 \\
  & 0 & w \\
 \mathbb{1}_{2k-1} &  &  \\
\end{array}
\right) \:.
\end{equation}
The computation of the localized zero modes proceeds with the same steps done with $k=1$. The Higgs branch is now given by \textbf{k free hypermultiplets}. This result confirms what found with different method in \cite{Closset:2020scj}.

\be
\boxed{uv = z^{2k+1} + w^2 \qquad \longleftrightarrow \qquad \text{k free hypers}}
\ee
As noted in \cite{Collinucci:2021wty}, different choices of the Higgs field can be made, corresponding to inequivalent T-brane backgrounds (putting the $1$'s and $w$'s in different entries) that would generate a smaller spectrum, and therefore a lower-dimensional Higgs branch.

Let us examine a simple example of a geometry that admits three possible Higgs branches, depending on the choice of T-brane data. Take the hypersurface given by
\be
u v = z^5+w^2\,.
\ee
Again, this is a singular hypersurface that does not admit a crepant K\"ahler resolution. Nevertheless, it does admit a fiberwise reduction to IIA string theory with D6-branes, albeit with D6-branes that wrap singular Riemann surfaces. For the Higgs background, we see the following three possible choices, each giving rise to a different hypermultiplet spectrum:
\begin{center}
$\begin{array}{ccc}
\Phi_2=\left(
\begin{array}{ccccc}
 z & w & 0 & 0 & 0 \\
 0 & z & 1 & 0 & 0 \\
 0 & 0 & z & 1 & 0 \\
 0 & 0 & 0 & z & w \\
 1 & 0 & 0 & 0 & z \\
\end{array}
\right) & \Rightarrow & 2 \text{ free hypers} \\
 & & \\
 & & \\
\Phi_1 = \left(
\begin{array}{ccccc}
 z & w & 0 & 0 & 0 \\
 0 & z & w & 0 & 0 \\
 0 & 0 & z & 1 & 0 \\
 0 & 0 & 0 & z & 1 \\
 1 & 0 & 0 & 0 & z \\
\end{array}
\right) & \Rightarrow & 1 \text{ free hyper}  \\
 & & \\
 & & \\
\Phi_0 = \left(
\begin{array}{ccccc}
 z & 1 & 0 & 0 & 0 \\
 0 & z & 1 & 0 & 0 \\
 0 & 0 & z & 1 & 0 \\
 0 & 0 & 0 & z & 1 \\
 w^2 & 0 & 0 & 0 & z \\
\end{array}
\right) & \Rightarrow & 0 \text{ free hypers}  \\
\end{array}$

\end{center}
In all three cases, there is no discrete gauging, so we just have free hypermultiplets. We refer to these different choices as T-brane data, as they consist in inherently non-Abelian information that does not alter the M-theory geometry, but nevertheless has a severe impact on the effective physics. It would be interesting to pursue such examples and compare results with other available methods.

\subsection{Partially resolvable singularities}

We finally consider a class of threefolds that are a straightforward generalization~of~\eqref{non-resol-simpleone}:
\begin{equation}\label{EqNonresolResol}
uv=z(z^{2k+1}+ w^2)\:.
\end{equation}
These spaces can be \emph{partially} resolved by taking
\begin{align}
\bm u & z\\z^{2k+1}+ w^2 & v \eem \cdot \bm s \\t \eem &= 0 \,,
\end{align}
where $[s:t]$ are the homogeneous coordinates of the exceptional $\pp^1$. However, the 
resolution still possesses a terminal singularity. Hence one expects hypers coming from M2-branes wrapping the exceptional $\pp^1$ and hypers coming from the terminal singularity. Moreover, we expect a continuous $U(1)$ flavor symmetry from the non-Cartier divisor related to the small resolution (like in all the previous cases where a small resolution was possible).

Again we study in detail only the case $k=1$. 
The Higgs field whose characteristic polynomial reproduces \eqref{EqNonresolResol} is 
\begin{equation}\label{higgs}
\Phi =\left(
\begin{array}{cccc}
 0 &  &  &  \\
  & 0 & w & 0 \\
  & 0 & 0 & w \\
  & 1 & 0 & 0 \\
\end{array}
\right) \:.
\end{equation}
The Higgs field fluctuations $\varphi$ are given modulo linearized gauge transformation:
\begin{equation}
\varphi \sim \varphi +[\Phi,g] \qquad \mbox{with} \qquad g \in \mathfrak{gl}(4)\:.
\end{equation}
Explicitly we get:
$$
\varphi \sim \underbrace{\left(
\begin{array}{cccc}
 \varphi_{11} & \varphi_{12}  & \varphi_{13}  & \varphi_{14}  \\
 \varphi_{21}  & \varphi_{22}  & \varphi_{23}  & \varphi_{24}  \\
 \varphi_{31}  & \varphi_{32}  & \varphi_{33}  & \varphi_{34}  \\
 \varphi_{41}  & \varphi_{42}  & \varphi_{43}  & \varphi_{44}  \\
\end{array}
\right)}_{\varphi} +\underbrace{\left(
\begin{array}{cccc}
 0 & g_{14} & g_{12} w & g_{13} w \\
 -g_{31} w & g_{24}-g_{32} w & (g_{22}-g_{33}) w & (g_{23}-g_{34}) w \\
 -g_{41} w & g_{34}-g_{42} w & (g_{32}-g_{43}) w & (g_{33}-g_{44}) w \\
 -g_{21} & g_{44}-g_{22} & g_{42} w-g_{23} & g_{43} w-g_{24} \\
\end{array}
\right)}_{[\Phi,g]}.
$$
Using the gauge redundancy we can  set $\varphi_{12}$, $\varphi_{41}$, $\varphi_{42}$, $\varphi_{43}$, $\varphi_{44}$ and $\varphi_{32}$ to zero.\\
We are then left with:
$$
\varphi \sim \underbrace{\left(
\begin{array}{cccc}
 \varphi_{11} & 0  & \varphi_{13}  & \varphi_{14}  \\
 \varphi_{21}  & \varphi_{22}  & \varphi_{23}  & \varphi_{24}  \\
 \varphi_{31}  & 0  & \varphi_{33}  & \varphi_{34}  \\
 0  & 0 & 0 & 0  \\
\end{array}
\right)}_{\varphi} +\underbrace{\left(
\begin{array}{cccc}
 0 & 0 & g_{12} w & g_{13} w \\
-g_{31}w & (g_{43} -g_{32})w & (g_{44}-g_{33}) w & 0 \\
 -g_{41}w & 0 & -(g_{43}-g_{32}) w & -(g_{44}-g_{33}) w \\
 0 & 0 & 0 & 0 \\
\end{array}
\right)}_{[\Phi,g]}.
$$
We see  that $\varphi_{11}$ and $\varphi_{24}$ are unconstrained, and that the pairs $(\varphi_{22}, \varphi_{33})$ and $(\varphi_{23}, \varphi_{34})$ depend on the same parameters, so that we can gauge-fix only a linear combination for each pair.
As a result (making a choice for the gauge-fixing), in total we get \textbf{6 modes}, or equivalently \textbf{3 hypers}:
\begin{equation}
 \varphi_{13},\varphi_{14},\varphi_{21},\varphi_{31} \in \mathbb{C}[w]/(w) \cong \mathbb{C} \qquad\mbox{and} \qquad
 \varphi_{22},\varphi_{23} \in \mathbb{C}[w]/(w) \cong \mathbb{C} 
\end{equation}

The subgroup of $SU(4)$ preserving $\Phi$ as in (\ref{higgs}) is given by matrices
\begin{equation}
G= \left(
\begin{array}{cc}
 e^{-3i \alpha } & 0  \\
 0 & e^{i \alpha }  \mathbb{1}_3\\
\end{array}
\right) .
%\hspace{1cm} \text{with } \alpha+3\beta =0 \text{ mod } 2\pi\:.
\end{equation}
A direct computation shows that the modes $ \varphi_{13},\varphi_{14},\varphi_{21},\varphi_{31}$ are charged under the $U(1)$, whereas $\varphi_{22},\varphi_{23}$ are not.
Summing up, we get \textbf{2 charged hypers} and \textbf{1 uncharged hyper.}

In the generic $k$ case we still have a $U(1)$ flavor and the modes are organized (as in $k=1$ case) as follows:
\begin{equation}
\left(
\begin{array}{c|c}
0 & \text{charged}_{1\times (2k+1)} \\
\hline
\text{charged}_{(2k+1)\times 1} &\text{uncharged}_{(2k+1)\times (2k+1)} \
\end{array}
\right)
\end{equation}
We obtain \textbf{2 charged hypers} along with $\boldsymbol{k}$ \textbf{uncharged hypers}. Since there is no discrete gauge symmetry, the Higgs Branch is given by $k+2$ free hypermultiplets.
\be
\boxed{uv=z(z^{2k+1}+ w^2)\:. \qquad \longleftrightarrow \qquad \text{2 charged hypers $+$ k uncharged hypers}}
\ee

\section{Conclusions}
In this work we introduced a new way of studying the Higgs branches of rank zero theories obtained from M-theory on CY threefolds that admit flops, and are either $\cc^*$-fibered, or $\zz_2$-orbifolds of such fibrations. Our techniques are based on the M/IIA duality, and readily reveal discrete gauge groups. The novelty of our approach is that it allows us to leave the realm of toric examples. It also allows us to study cases that do not admit crepant resolutions.

Our results can be generalized to theories that admit more complicated flop structures. More interestingly, such rank zero theories could be coupled to higher rank theories, in which case the Higgs branch structure will be richer.

It would be important to be able to reproduce our results purely from the monodromy data of complex structure deformations of the threefolds, and Wilson lines of the $C_3$-field. At the very least, it should be possible to connect that M-theory data to the deformation theory of D6-branes on Riemann surfaces and the corresponding Wilson lines of the DBI gauge fields. 
This is rather convoluted, and has not even been done in a clear cut way for the \emph{noncompact} conifold, despite the pioneering work of \cite{Ooguri:1996me}.

%%%%%%%%%%%%%%%%%%%%%%%%%%%%%%%%%%%%%%%%%%%%%%%%
\section*{Acknowledgments} % sec (acknow)
%%%%%%%%%%%%%%%%%%%%%%%%%%%%%%%%%%%%%%%%%%%%%%%%

We have benefited from discussions with A. Bourget, C. Closset and S. Sch\"afer-Nameki. A.C.~is a Research Associate of the Fonds de la Recherche Scientifique F.N.R.S.~(Belgium). The work of A.C.~is partially supported by IISN - Belgium (convention 4.4503.15), and supported by the Fonds de la Recherche Scientifique - F.N.R.S.~under Grant CDR J.0181.18. 
The work of R.V.~is partially supported by ``Fondo per la Ricerca di Ateneo - FRA 2018'' (UniTS). 
A.S. and R.V. acknowledge support by INFN Iniziativa Specifica ST\&FI. M.D.M. acknowledges support by INFN Iniziativa Specifica GAST.

% fold sec (acknow)

%%%%%%%%%%%%%%%%%%%%%%%%%%%%%%%%%%%%%%%%%%%%%%%%

%%%%%%%%%%%%%%%%%%%%%%%%%%
% BIBLIOGRAPHY
%%%%%%%%%%%%%%%%%%%%%%%%%%

\providecommand{\href}[2]{#2}

\end{document}